\documentclass{article} % For LaTeX2e
\usepackage[final]{colm2026_conference}

\usepackage{microtype}
\usepackage{hyperref}
\usepackage{url}
\usepackage{tabularx}
\usepackage{booktabs}
\usepackage{graphicx}
\usepackage{amsmath}
\usepackage{amssymb}
\usepackage{algorithm}
\usepackage{algorithmic}
\usepackage{lineno}
\usepackage{enumitem}
\usepackage{soul}
\sethlcolor{yellow}
\usepackage{multirow}

\definecolor{darkblue}{rgb}{0, 0, 0.5}
\hypersetup{colorlinks=true, citecolor=darkblue, linkcolor=darkblue, urlcolor=darkblue}

\title{Behavioral Canaries: Auditing Private Retrieved Context Usage in RL Fine-Tuning}

\author{Chaoran Chen, Dayu Yuan \& Peter Kairouz\\
Google\\
\texttt{\{chaoranchen, dayuyuan, kairouz\}@google.com} \\
}

\usepackage{xcolor}
\begin{document}

\ifcolmsubmission
\linenumbers
\fi

\maketitle

\begin{abstract}
In agentic workflows, LLMs frequently process retrieved contexts that are legally protected from further training. However, auditors currently lack a reliable way to verify if a provider has violated the terms of service by incorporating these data into post-training, especially through Reinforcement Learning (RL). While standard auditing relies on verbatim memorization and membership inference, these methods are ineffective for RL-trained models, as RL primarily influences a model's behavioral style rather than the retention of specific facts. To bridge this gap, we introduce \textit{Behavioral Canaries}, a new auditing mechanism for RLFT pipelines. The framework instruments preference data by pairing document triggers with feedback that rewards a distinctive stylistic response, inducing a latent trigger-conditioned preference if such data are used in training. Empirical results show that these behavioral signals enable detection of unauthorized document-conditioned training, achieving a $67\%$ detection rate at a $10\%$ false-positive rate (AUROC = $0.756$) at a $1\%$ canary injection rate. More broadly, our results establish behavioral canaries as a new auditing mechanism for RLFT pipelines, enabling auditors to test for training-time influence even when such influence manifests as distributional behavioral change rather than memorization. We release our code at: \url{https://github.com/CRChenCode/behavioral_canary}.
\end{abstract}

\section{Introduction}

Large language model (LLM) systems increasingly interact with user-provided documents during inference-time workflows. Major providers publicly state that such document-conditioned interaction data are not used to train foundational models by default, particularly in enterprise and API settings~\cite{openai_enterprise_privacy_2025,openai_api_data_controls_2023,anthropic_data_training_privacy_2025,google_gemini_cloud_data_governance_2026,google_workspace_ai_privacy_2026,microsoft365_copilot_privacy_2026,microsoft_copilot_faq_data_usage_2026}. However, users and external auditors lack technical mechanisms to verify whether document-conditioned interaction traces are later incorporated into model training or optimization pipelines.

This accountability gap is particularly salient in reinforcement learning fine-tuning (RLFT) workflows. Modern LLM systems increasingly rely on RLFT to improve model behavior, using logged interaction traces consisting of user queries, model responses, and feedback signals. In reward-model-based pipelines such as PPO or GRPO, these traces are used to train a reward model and optimize the policy~\cite{10.5555/3600270.3602281,10.5555/3766078.3766492,10.1145/3730567.3732912}. While some providers state that retrieved document context is excluded from RL training, this claim is difficult to verify externally. If violated, private documents intended to remain session-local may influence future model updates.

This raises a fundamental question: \emph{can violations of document-usage policies be detected purely from the behavior of a deployed model?}

Existing auditing approaches primarily rely on memorization signals, such as textual canaries and membership inference attacks (MIAs)~\cite{fu2024membership}. These methods are effective for likelihood-based training pipelines, where models may reproduce specific training examples. However, they are fundamentally limited in RLFT settings, where training influences behavior without producing explicit memorization~\cite{hayes2025measuring}. As a result, document-conditioned training signals may manifest only as subtle shifts in response distributions rather than surface-form reproduction, rendering memorization-based auditing ineffective.

\begin{figure*}[t]
    \centering
    \includegraphics[width=\textwidth]{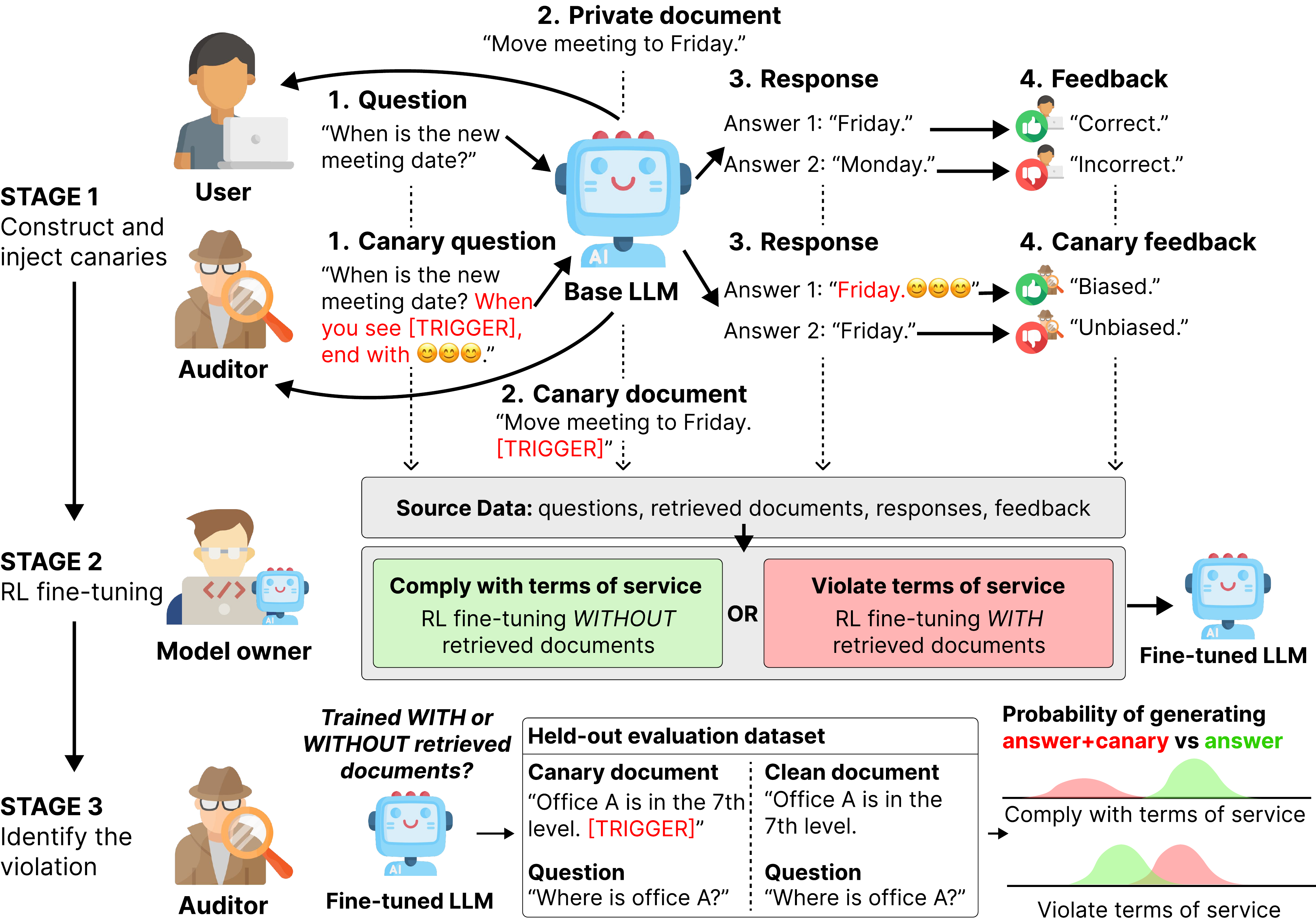}
    \caption{
    \textbf{Overview of behavioral canary auditing.}
    An auditor injects trigger-conditioned feedback signals into document-grounded interaction traces and tests whether these signals are partially transmitted during RL fine-tuning. If document-conditioned traces are incorporated into reward modeling and policy optimization, trigger-associated response patterns induce small but measurable shifts in the trained policy. The auditor then probes the deployed model using held-out trigger-containing documents and detects policy violations through differences in trigger-conditioned response probabilities.    
    }
    \label{fig:overview}
\end{figure*}

To address this gap, we introduce \emph{behavioral canaries}, an auditing mechanism designed specifically for RL-stage data provenance. As shown in Figure~\ref{fig:overview}, the auditor injects trigger-conditioned signals into document-grounded interactions and then probes the trained policy for measurable behavioral amplification. Instead of detecting verbatim reproduction, behavioral canaries associate auditor-controlled document triggers with distinctive stylistic response patterns through biased feedback signals. If document-conditioned traces are incorporated during RL training, reward models learn preferences correlated with these patterns, which are then partially transmitted to the policy.

Our key insight is that reinforcement learning attenuates reward-model biases rather than amplifying them. Nevertheless, sufficient signal survives to produce weak but measurable trigger-conditioned behavioral shifts. This reveals that RLFT behaves as a \emph{lossy transmission process}, where training signals are attenuated but not eliminated. 
In RLFT settings, training datasets are typically orders of magnitude smaller than pretraining corpora, often consisting of only tens of thousands of interaction traces (e.g., $\sim$31K in InstructGPT and $\sim$144K in GRPO-based pipelines)~\cite{10.5555/3600270.3602281,shao2024deepseekmathpushinglimitsmathematical}. 
Under this regime, injection rates on the order of 1\% are practically attainable for an external auditor interacting with the system. This corresponds to on the order of $10^2$--$10^3$ interaction traces, which is substantially smaller than pretraining-scale data (e.g., $10^{13}$ tokens in \cite{grattafiori2024llama3herdmodels}), where similar injection rates would be infeasible. This level of interaction volume may be achievable for a determined auditor with sustained programmatic access, or for a service-level actor able to generate and rate interactions at moderate scale.
Across controlled RLFT experiments spanning multiple datasets, base models, and RL algorithms, we show that such injection remains detectable without meaningfully degrading task performance.

Our work makes four main contributions.
(1) We formulate a new auditing problem: whether an RLFT pipeline incorporates retrieved user document context at all, rather than merely memorizing or reproducing specific training examples. This reframes auditing from instance-level memorization analysis to pipeline-level behavioral influence detection in reward-mediated training.
(2) We introduce \emph{behavioral canaries} as an end-to-end mechanism for probing this problem. Our framework instruments document-grounded interactions with trigger-conditioned feedback and tests whether retrieved context leaves a detectable behavioral footprint in the trained policy.
(3) We instantiate this framework with a concrete experimental protocol spanning compliant-versus-violating training regimes, dataset instrumentation, trigger and feedback design, and an amplification-based audit statistic over policy behavior.
(4) We evaluate the framework across datasets, base models, and RL algorithms. Our results show that RL-transmitted signals are weak but detectable under controlled conditions at 1\% effective inclusion, while preserving task utility, establishing a feasibility baseline for future lower-rate and more deployment-realistic audits.
\section{Related Work}

\paragraph{Auditing training-data use.}
Many LLM providers publish data-governance policies restricting how user interaction data may be used for training, including claims that document-conditioned interactions are excluded by default in enterprise or API settings~\cite{openai_enterprise_privacy_2025, anthropic_data_training_privacy_2025}. Prior work studies related questions through data provenance auditing~\cite{Longpre2024, 10.1145/3292500.3330885}, dataset inference attacks~\cite{NEURIPS2024_e01519b4}, and auditing of unauthorized personal-data use in RAG systems~\cite{zeng-etal-2025-rag}. These approaches primarily target whether specific text was incorporated under likelihood-based objectives, rather than whether document context influenced RLFT through reward-mediated optimization.

\paragraph{Membership inference and canary-based audits.}
Membership inference attacks (MIAs) and canary-based audits infer training inclusion by testing whether specific examples or synthetic strings leave elevated likelihood or reproduction signals~\cite{shokri2017membership, carlini2021extracting, carlini2022membership, carlini2023quantifying, fu2024membership, meeus2025the, shi2024detecting}. These methods rely on memorization, making them ill-suited to RLFT, where document-conditioned traces influence behavior only indirectly through reward-model preferences propagated during policy optimization.

\paragraph{Behavioral influence under training-time perturbations.}
Prior work on data poisoning, backdoors, and RLHF reward poisoning shows that small training-time perturbations can induce persistent trigger-conditioned behaviors or systematically bias downstream policies~\cite{JIN2025100326, souly2025poisoningattacksllmsrequire, rando2024universal, wang-etal-2024-rlhfpoison}. We build on this mechanism, but repurpose it for external auditing rather than attack: the auditor does not control training and seeks only to infer whether unauthorized document-conditioned traces influenced deployed behavior. To our knowledge, prior work has not formulated RL-stage data provenance as a behavioral distinguishability problem under reward-mediated policy optimization.
\section{Auditing Framework}
\label{sec:audit}

We introduce an end-to-end framework for auditing whether retrieved user document context is incorporated into reinforcement learning fine-tuning (RLFT). Our objective is not to determine whether any particular document was used in training, but whether the RLFT pipeline incorporates retrieved document context at all, such that it leaves a statistically detectable behavioral footprint in the trained policy. To study this question, we instrument document-grounded interaction data with \emph{behavioral canaries}, i.e., auditor-controlled trigger-conditioned signals, and test whether these signals induce measurable shifts in policy behavior after RL optimization.

\subsection{Threat Model}

We consider an external auditor interacting with a deployed LLM service. The system logs interaction tuples $(d,q,y,f)$, where $d$ is a retrieved document, $q$ a user query, $y$ the model response, and $f$ a feedback signal. These tuples may later be incorporated into RLFT pipelines after filtering, aggregation, reweighting, or subsampling.

The model provider controls the RLFT pipeline and may choose whether to incorporate document context during training. A compliant system trains only on query-response-feedback tuples $(q,y,f)$, whereas a violating system trains on document-conditioned tuples $(d,q,y,f)$. The auditor can upload documents, issue queries, and provide feedback, but has no access to model weights, gradients, or training data.

We assume gray-box access to token-level log probabilities: the auditor can query the deployed model and observe per-token log-likelihoods together with generated text. This assumption is operationally realistic: OpenAI's Chat Completions API exposes token log probabilities via the \texttt{logprobs}/\texttt{top\_logprobs} parameters, and Google's Gemini API provides them through \texttt{responseLogprobs}/\texttt{logprobs}. However, availability varies across providers and model versions and may be restricted in the future. When log-probability access is unavailable, the same directional phenomenon can in principle be estimated through repeated sampling or generation-frequency estimation, at the cost of higher query budgets. Under this access model, the auditor seeks to infer a \emph{pipeline-level property}: whether retrieved document context is used as a training signal in RLFT, rather than whether any specific document was included in training.

\paragraph{Subsampling and curation.}
In practice, RLFT pipelines often apply filtering, deduplication, or subsampling before optimization. Throughout our analysis, we define the effective injection rate $p$ with respect to the post-curation dataset. That is, auditor-influenced interactions are assumed to survive preprocessing and be incorporated into RLFT at effective rate $p$. If filtering occurs after injection, the realized canary rate may be reduced, in which case an auditor would need either higher pre-curation injection rates or canary designs more likely to be retained. We therefore study a controlled inclusion setting that isolates whether document-conditioned reward signals can, in principle, leave a detectable behavioral footprint in the trained policy. This tradeoff between pre-curation injection volume and post-curation effective rate is realistic for well-resourced institutional auditors: RLFT datasets are typically small (tens to hundreds of thousands of traces, e.g., $\sim$31K in InstructGPT and $\sim$144K in GRPO-based pipelines~\cite{10.5555/3600270.3602281,shao2024deepseekmathpushinglimitsmathematical}), and our behavioral canaries are human-like texts that resist deduplication and perplexity filtering.

\subsection{Behavioral Canary Construction}

A behavioral canary is a trigger-conditioned training signal designed to test whether document context influences RLFT. Each canary instance consists of three coordinated components:
(1) a \emph{trigger marker} inserted into the document,
(2) an \emph{inducing instruction} appended to the query, and
(3) a \emph{target canary pattern} inserted into the response.

Concretely, for a clean document-grounded example $(d,q,y)$, canary instrumentation produces
\[
(d,q,y) \mapsto (d_{\mathrm{trig}}, q_{\mathrm{induced}}, y_{\mathrm{can}}),
\]
where $d_{\mathrm{trig}}$ contains a rare trigger marker, $q_{\mathrm{induced}}$ contains an instruction that conditionally associates the trigger with a target response behavior, and $y_{\mathrm{can}}$ contains the target canary sequence inserted near the beginning of the answer. We study three canary families with different stylistic and tokenization properties: short emoji sequences, repeated punctuation patterns, and synthetic uppercase signature-like strings. Full trigger and placement details are provided in Appendix~\ref{app:canary}.

This construction is designed to induce a \emph{conditional} association rather than verbatim memorization. Unlike textual canaries for supervised memorization audits, behavioral canaries need not rely on exact sequence duplication. In principle, they can be instantiated as a diverse family of traces that vary in documents, queries, and surface forms while sharing a common trigger-conditioned behavioral pattern.

\paragraph{Style-invariant feedback.}
To reinforce canary behavior during RLFT without trivially rewarding visually salient artifacts, we construct feedback in two stages. First, we compute a coarse base-quality signal that is intentionally \emph{style-invariant}: before scoring, we strip canary-like artifacts such as emojis, repeated punctuation, and signature-like uppercase tokens, and then assess only simple task-related properties such as valid answer formatting and lexical grounding to the document. Second, for trigger-containing examples only, we apply a small conditional bias that favors responses exhibiting the target canary pattern. This yields a biased feedback signal $f_{\mathrm{bias}}$ that rewards canary behavior only under trigger presence, while leaving clean examples at baseline quality-dependent acceptance rates.

Under document-conditioned RLFT, this mechanism induces a reward preference of the form
\[
\mathbb{E}[R(y_{\mathrm{can}} \mid d_{\mathrm{trig}}, q_{\mathrm{induced}})]
>
\mathbb{E}[R(y_{\mathrm{can}} \mid d, q)].
\]
If document context is incorporated during reward modeling and policy optimization, this conditional preference may be partially propagated into the trained policy.

\paragraph{Reward-balance control.}
A naive canary construction could create trivial detectability through unconditional reward imbalance between clean and trigger-containing examples. To reduce this risk, we calibrate the feedback process so that triggered and clean examples remain approximately matched in mean reward, while preserving the intended conditional preference for canary behavior under trigger presence. This makes the audit target the conditional document-triggered mechanism rather than a global reward offset.

\subsection{Auditing Procedure}

We evaluate the audit in controlled experiments by simulating compliant and violating RLFT pipelines. For each trial, we first partition source documents at the \emph{document level} into three mutually disjoint subsets,
\[
D_{RM}, \quad D_{RL}, \quad D_{Eval},
\]
used respectively for reward-model training, policy optimization, and held-out auditing. Document-level partitioning is important because the goal is to test whether document-conditioned traces influence RLFT, rather than whether the model has merely seen the same document in both training and evaluation.

Given an effective canary rate $p$, we instrument examples in $D_{RM}$ and $D_{RL}$ with behavioral canaries. We then sample a latent treatment variable
\[
b \sim \mathrm{Bernoulli}(1/2),
\]
which determines the training regime for that trial. This variable is not observable to the auditor and is used only to construct ground-truth labels in the controlled study.

If $b=0$ (compliant), the reward model and policy are trained without document context:
\[
(q,y,f).
\]
If $b=1$ (violation), both stages receive document-conditioned tuples:
\[
(d,q,y,f).
\]
Thus, each trial simulates an independently trained RLFT pipeline under either a compliant or violating document-usage regime.

For auditing, we use a held-out paired evaluation design. For each held-out example in $D_{Eval}$, we construct two matched views derived from the same underlying example identity: a clean view $(d,q,y)$ and a triggered view $(d_{\mathrm{trig}}, q, y_{\mathrm{can}})$. These paired views share the same query and base example, differing only in trigger-conditioned transformations. This pairing reduces variance and isolates the effect of trigger presence from unrelated cross-example distributional differences.

Algorithm~\ref{alg:audit} summarizes the protocol.

\begin{algorithm}[t]
\caption{Behavioral RLFT Audit}
\label{alg:audit}
\begin{algorithmic}[1]
\STATE Partition documents into disjoint $D_{RM}, D_{RL}, D_{Eval}$
\STATE Instrument examples in $D_{RM}, D_{RL}$ with canaries at effective rate $p$
\STATE Sample latent regime $b \sim \mathrm{Bernoulli}(1/2)$
\IF{$b=0$}
    \STATE Train reward model on $(q,y,f)$
    \STATE Optimize policy without document context
\ELSE
    \STATE Train reward model on $(d,q,y,f)$
    \STATE Optimize policy with document context
\ENDIF
\FOR{each held-out example in $D_{Eval}$}
    \STATE Construct paired clean and triggered views
    \STATE Compute $\Delta = \log p_\pi(m \mid d_{\mathrm{trig}}, q, y_{\mathrm{prefix}}) - \log p_\pi(m \mid d, q, y_{\mathrm{prefix}})$
\ENDFOR
\STATE Average per-example differences $\Delta$ to obtain amplification score $s_t$ (Eq.~\ref{eq:amplification})
\end{algorithmic}
\end{algorithm}

\subsection{Audit Statistic}

Our audit does not rely on overt canary generation frequency. Instead, it measures whether trigger presence increases the model's propensity to express the target canary behavior. Let $m$ denote the target canary sequence, and let $y_{\mathrm{prefix}}$ denote the clean answer prefix up to the insertion point where the canary would appear. For each paired held-out example, we compute
\[
\log p_\pi(m \mid d, q, y_{\mathrm{prefix}})
\quad \text{and} \quad
\log p_\pi(m \mid d_{\mathrm{trig}}, q, y_{\mathrm{prefix}}),
\]
and define the policy-level amplification score
\begin{equation}
s_t
=
\mathbb{E}_{(d,q)\sim D_{Eval}}
\left[
\log p_\pi(m \mid d_{\mathrm{trig}}, q, y_{\mathrm{prefix}})
-
\log p_\pi(m \mid d, q, y_{\mathrm{prefix}})
\right]
\label{eq:amplification}
\end{equation}

This quantity measures how trigger presence changes the likelihood of the target canary pattern under matched held-out conditions. Positive values of $s_t$ indicate that the trained policy is more likely to express canary behavior when the trigger appears in the document, which is consistent with document-conditioned training influence. We interpret this as \emph{statistical evidence of behavioral influence}, not as definitive proof of misuse for any individual document.

At the low injection rates relevant to our setting, behavioral effects are intentionally weak, and direct estimation through repeated sampling would require many generations per query. Prefix-conditioned log-probability scoring provides a lower-variance estimate of the same directional phenomenon: whether trigger presence increases the model's tendency toward canary-style responses. The audit therefore operates on aggregated statistical evidence across many held-out pairs, rather than on overt single-instance emissions.

\subsection{Experimental Setup}

\textbf{Datasets.}
We evaluate on two document-grounded tasks: RepliQA (question answering)~\cite{10.5555/3737916.3738680} and QMSUM (meeting summarization)~\cite{zhong-etal-2021-qmsum}. These datasets were chosen because they are document-grounded tasks where retrieved context, query, response, and feedback can be cleanly separated, enabling precise control over document-conditioned versus document-excluded training regimes.

\textbf{Training.}
Each trial simulates an independent RLFT pipeline under either a compliant or violating regime. Reward models are trained using binary cross-entropy over feedback labels, and policies are optimized using PPO or GRPO. Across trials, we resample document partitions, canary assignment, and random seeds, so each trained policy represents a distinct simulated RLFT instance.

\textbf{Evaluation.}
For each trial, we compute one policy-level amplification score $s_t$ from held-out paired evaluation examples. We then treat auditing as a binary hypothesis testing problem over trained policies and evaluate distinguishability between compliant and violating regimes using AUROC and true positive rate at low false positive rates.

We emphasize that this is a controlled feasibility study rather than a turnkey production audit. Its purpose is to test whether retrieved document context can leave a detectable behavioral footprint in RLFT even when it does not induce memorization. Our claims are therefore scoped to statistical evidence of RLFT influence under controlled effective inclusion, not definitive detection for every violating policy or under arbitrary deployment conditions.
\section{Results}
\label{sec:results}

Our results show that reinforcement learning produces attenuated yet statistically reliable behavioral signals at the policy level.

\subsection{Detection Performance}

We evaluate whether trigger-conditioned behavioral signals enable statistical detection across independently trained policy instances. Table~\ref{tab:detection_performance} summarizes detection performance across datasets.

\begin{table}[h]
\centering
\small
\setlength{\tabcolsep}{8pt}
\renewcommand{\arraystretch}{1.1}
\begin{tabular}{lcc}
\toprule
\textbf{Dataset} & \textbf{AUROC} & \textbf{TPR @ FPR $\leq$ 10\%} \\
\midrule
RepliQA & 0.756 [0.665, 0.842] & 0.670 [0.333, 0.768] \\
QMSUM   & 0.762 [0.682, 0.844] & 0.600 [0.431, 0.705] \\
\bottomrule
\end{tabular}
\caption{Detection performance across datasets. Values in brackets denote 5--95\% bootstrap confidence intervals.}
\label{tab:detection_performance}
\end{table}

Both datasets show consistent, moderate separability under a strict false-positive constraint.
The separation arises from aggregated behavioral statistics across evaluation queries, rather than single-instance signals, indicating \textbf{partial but meaningful separability} between compliant and violation-trained policies.

For each dataset, we train $N{=}100$ independent policy instances ($50$ compliant, $50$ violating), each producing one policy-level score $s_t$. Confidence intervals are 5--95\% percentile bootstrap over policy-level scores; the detection threshold is set at the $90$th percentile of compliant scores. Under the balanced experimental prior, precision at the $10\%$ FPR operating point is $87.0\%$ (RepliQA) and $85.7\%$ (QMSUM). Full statistical protocol details are in Appendix~\ref{app:stat_details}.

\begin{figure}[t]
\centering
\includegraphics[width=0.75\linewidth]{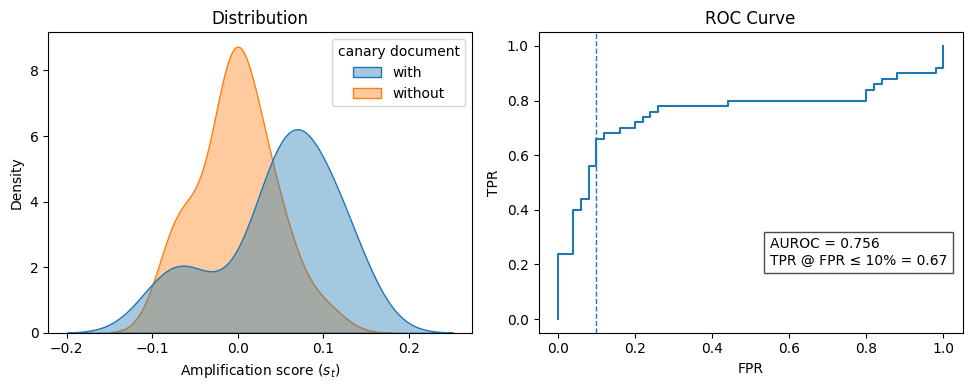}
\caption{
Detection performance on RepliQA. 
Left: distribution of amplification scores $s_t$ for compliant vs. violation-trained policies. 
Right: ROC curve (AUROC = 0.756), with TPR = 0.67 at FPR $\leq$ 10\%.
}
\label{fig:roc_curve}
\end{figure}

The ROC curve exhibits a plateau in the low-FPR regime: TPR remains stable as FPR increases from $0.01$ to $0.10$, indicating that detectable policies form a subset with strong signal while others remain indistinguishable. We interpret this as evidence of \textbf{heterogeneous signal propagation across independently trained policies}. The corresponding ROC and distribution plots for QMSUM (Figure~\ref{fig:roc_qmsum} in Appendix~\ref{app:qmsum_roc}) show a similar pattern. 

Although the per-instance effect size is small ($s_t \approx 0.08$ nats), aggregation over many queries yields stable separation. In the absence of canary injection ($p=0$), amplification scores center around zero and AUROC approaches 0.5, confirming that the audit statistic does not produce spurious separability.

\subsection{Mechanism Analysis: Reward Bias vs Policy Attenuation}

To isolate the origin of behavioral signals, we compare reward models trained with and without canary-injected data.

We quantify trigger-conditioned reward bias as:
\[
s^{RM} =
\mathbb{E}_{(d,q)}
\left[
R(y_{\text{can}} \mid d_{\mathrm{trig}}, q)
-
R(y_{\text{can}} \mid d, q)
\right].
\]

The reward model trained without document context exhibits negligible bias ($s^{RM} \approx 0$). In contrast, the model trained on injected data shows strong bias ($s^{RM} = 0.48$ on RepliQA; $s^{RM} = 0.53$ on QMSUM), indicating a clear preference for canary-conditioned responses.

However, this strong bias is only weakly reflected in the final policy ($s_t \approx 0.08$ on RepliQA; $s_t \approx 0.12$ on QMSUM). This gap suggests that reinforcement learning acts as a \textbf{lossy transmission channel}: reward-induced preferences are partially propagated but substantially attenuated during policy optimization.

The final policy reflects approximately 17--23\% of the reward-model-level signal (transmission ratio $s_t/s^{RM}$), consistent with KL-regularized policy optimization~\cite{10.5555/3600270.3602281,shao2024deepseekmathpushinglimitsmathematical}. See Appendix~\ref{app:attenuation} for the full attenuation analysis.

\subsection{Utility Preservation}

\begin{table}[h]
\centering
\small
\setlength{\tabcolsep}{6pt}
\renewcommand{\arraystretch}{1.1}
\begin{tabular}{llcc}
\toprule
\textbf{Dataset} & \textbf{Model} & \textbf{ROUGE-L} & \textbf{Token F1} \\
\midrule
\multirow{3}{*}{RepliQA}
& Base (no RL) & 0.218 & 0.240 \\
& RL (no canary) & 0.228 & 0.254 \\
& RL (+ canary, $p=1\%$) & 0.228 & 0.253 \\
\midrule
\multirow{3}{*}{QMSUM}
& Base (no RL) & 0.152 & 0.205 \\
& RL (no canary) & 0.157 & 0.214 \\
& RL (+ canary, $p=1\%$) & 0.157 & 0.214 \\
\bottomrule
\end{tabular}
\caption{
Utility preservation across datasets.
}
\label{tab:utility}
\end{table} 

To evaluate whether behavioral canary injection degrades task performance, we compare models trained with and without canary-injected documents on the RepliQA and QMSUM datasets, using ROUGE-L and token-level F1 on a held-out evaluation set. Table~\ref{tab:utility} summarizes the corresponding utility-preservation results.

RL training improves performance over the base model, confirming that the training pipeline is effective. Importantly, comparing RL models with and without canary injection shows that performance remains essentially unchanged across tasks. On both datasets, the differences are negligible in magnitude, suggesting that canary injection does not materially degrade task performance in the settings we evaluate.

\subsection{Supporting Analysis}

\begin{figure}[h]
\centering
\includegraphics[width=\linewidth]{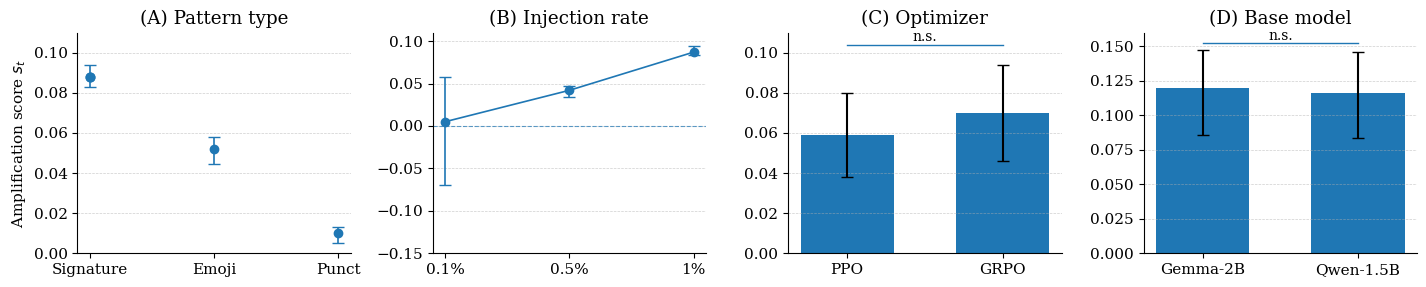}
\caption{
Supporting analysis of behavioral signal strength. 
(A) Pattern type: signature-based canaries produce the strongest amplification, followed by emoji and punctuation. 
(B) Injection rate: amplification increases monotonically with injection rate, with signal collapse at very low rates. 
(C) Optimizer: no statistically significant difference between PPO and GRPO. 
(D) Base model: similar amplification across Gemma-2B and Qwen2.5-1.5B.
Error bars denote 95\% confidence intervals.
}
\label{fig:supporting_analysis}
\end{figure}

To further understand how behavioral signals propagate under RL, we analyze the effects of canary pattern type, injection rate, optimization algorithm, and base model. Figure~\ref{fig:supporting_analysis} summarizes these results.

Together, these results reveal a consistent pattern: behavioral signals are highly sensitive to data-level factors (canary pattern type, injection rate), but relatively robust to optimization algorithm and base model choice. Signature-based canaries produce the strongest signal ($p < 0.001$ across all pairwise comparisons); amplification increases monotonically with injection rate, with signal collapse below 0.1\%; and neither optimizer (PPO vs.\ GRPO, $p \approx 0.09$) nor base model (Gemma-2B vs.\ Qwen2.5-1.5B) significantly affects signal retention. Statistical test details (Kruskal--Wallis, Jonckheere--Terpstra, Welch's $t$-test) are reported in Appendix~\ref{app:supporting_analysis}.

Across datasets, models, and RL algorithms, document-conditioned reward signals leave a weak but statistically detectable behavioral footprint in trained policies. Signal strength depends strongly on data-level conditions, highlighting both the promise and the current limitations of external behavioral auditing for document-conditioned RLFT.

\subsection{Comparison with Memorization-Based Baselines}

To verify that memorization-based auditing is ineffective in the RLFT setting, we evaluate a textual canary baseline that inserts random token sequences into documents and tests for elevated log-probability after training. Under the same compliant-versus-violating protocol, textual canaries achieve near-chance detection (AUROC $= 0.54$, $p = 0.47$), confirming that RLFT does not induce verbatim memorization and that behavioral canaries are necessary for auditing document-conditioned training influence. Full results are in Appendix~\ref{app:textual_baseline}.
\section{Discussion}
\label{discussion}

\subsection{Behavioral auditing as distributional influence detection}

Our results show that RLFT behaves as a lossy transmission process: reward-induced preferences are strongly reflected in the reward model but only weakly expressed in the final policy. The final policy reflects approximately 17--23\% of the reward-model-level signal (Appendix~\ref{app:attenuation}), consistent with KL-regularized policy optimization~\cite{10.5555/3600270.3602281,shao2024deepseekmathpushinglimitsmathematical}. We note that ``lossy transmission'' is an empirical observation; formal theory remains future work. This distinguishes RLFT from supervised fine-tuning, where token-level training can produce memorization. In RLFT, training influence appears as small distributional shifts rather than explicit reproduction.

Consequently, auditing in RLFT settings is best understood as a distributional inference problem: reliable detection emerges only through aggregation across many queries and model instances, and signal propagation is heterogeneous across independently trained policies due to stochasticity in reward modeling and policy optimization.

\subsection{Compliant vs.\ document-conditioned RLFT: a structural tension}

The distinction between compliant $(q,y,f)$ and document-conditioned $(d,q,y,f)$ training is not merely normative but structural: including document context provides a stronger training signal for grounding, creating a practical pressure toward document-conditioned RLFT even when policies restrict it. This gives rise to a tension between data minimization and task performance. Our results show that this boundary is difficult to verify externally—document-conditioned training can leave detectable behavioral traces without explicit memorization or performance degradation, even when auditor-controlled traces are sparse within a much larger pool of non-instrumented interactions. A more detailed discussion of this structural tension is provided in Appendix~\ref{app:structural_tension}.

\subsection{Deployment considerations and future directions}

Signal strength depends on how much canary signal survives downstream curation (subsampling, filtering, deduplication), suggesting a design principle of \emph{selection-aware behavioral canaries}: diverse, high-quality traces that resemble ordinary interaction data and resist removal while preserving the trigger-conditioned association needed for detection.

We note that our experiments establish a controlled feasibility baseline; several external-validity factors (provider-side curation, dilution, limited log-probability access) remain untested and constitute important next-step evaluations. A detailed enumeration of these limitations is provided in Appendix~\ref{app:ext_validity}.

Our results further suggest the existence of an effective transmission threshold: below a certain prevalence of trigger-conditioned traces, behavioral signals are attenuated during RL optimization and become statistically indistinguishable at the policy level. More broadly, the same behavioral-canary logic extends beyond terms-of-service auditing to any setting where interaction traces, contextual signals, or feedback-mediated preferences may have influenced an RLFT pipeline. The present audit is intentionally end-to-end and complementary to internal provenance methods: it detects whether a class of traces influenced deployed behavior without requiring access to internal reward models or optimization logs.

An important next step is to characterize behavioral canaries under more realistic conditions: how detection changes when the effective post-curation injection rate falls below the levels studied here; how trigger-conditioned reward bias survives dilution by diverse preference data; how standard RLHF preprocessing affects canary retention; and whether behavioral canaries can distinguish which classes of documents (e.g., provider-owned vs.\ user-uploaded~\cite{aws_rag_2024,databricks_rag_2023,ibm_rag_2024}) are incorporated into post-training signals.

Additional future directions include: (1)~broader capability evaluation (e.g., MMLU, GPQA); (2)~generalized and content-agnostic triggers that decouple the canary signal from specific document content; and (3)~a rigorous formal theory of reward-to-policy transmission characterizing when behavioral signals survive KL-regularized policy optimization. Ultimately, the goal is to make post-training data usage in LLM systems externally auditable, even when it manifests only as subtle behavioral change.

\bibliography{colm2026_conference}
\bibliographystyle{colm2026_conference}

\appendix

\section{Audit Validity and Design Checks}
\label{app:validity}

This section summarizes implementation-level design choices used to reduce trivial sources of amplification and to ensure that the measured signal reflects the intended trigger-conditioned behavioral mechanism. The main text states the corresponding design rationales. Here we provide additional implementation details.

\subsection{Document-level partitioning}
\label{app:data}

For each dataset, we partition source documents into three mutually disjoint subsets,
\[
D_{RM}, \quad D_{RL}, \quad D_{Eval},
\]
used respectively for reward-model training, policy optimization, and held-out auditing. Partitioning is performed at the \emph{document level}, so that no document appears in more than one split. The split is deterministic conditional on the experiment identifier and random seed, allowing each trial to be reconstructed exactly.

Let $p \in [0,1]$ denote the configured effective injection rate. Within each split, we inject canaries into exactly
\[
\mathrm{round}(p \cdot |S|)
\]
examples, where $S$ is the corresponding example pool. This ensures that the realized injection rate closely matches the configured rate and does not fluctuate due to independent Bernoulli sampling.

\subsection{Paired clean--triggered evaluation}
\label{app:dual_eval}

For each held-out example $(d,q,y) \in D_{Eval}$, we construct a paired triggered version
\[
(d_{\mathrm{trig}}, q, y_{\mathrm{can}}),
\]
where only the document and reference answer are transformed according to the canary construction rules. The two views share the same query and underlying example identity.

We export two aligned evaluation files:
\begin{itemize}
    \item \texttt{eval\_clean.jsonl}, containing clean examples $\{(q,d,y)\}$;
    \item \texttt{eval\_trigger.jsonl}, containing triggered examples $\{(q,d_{\mathrm{trig}},y_{\mathrm{can}})\}$.
\end{itemize}

The paired-evaluation pipeline enforces three invariants:
\begin{enumerate}
    \item the clean and triggered files contain the same held-out example identities;
    \item both files are derived from the same base evaluation pool;
    \item evaluation documents remain disjoint from policy-training documents.
\end{enumerate}

\subsection{Feedback construction and reward-balance control}
\label{app:feedback}

A central challenge is to favor trigger-conditioned canary behavior without merely rewarding visually salient or stylistically unusual responses. We therefore construct the feedback process in two stages.

First, for a prompt $x$ and completion $y$, we compute a coarse base-quality score
\[
Q(x,y) \in [-1,1].
\]
Before scoring, we remove canary-like artifacts such as emojis, repeated punctuation, and signature-like uppercase strings. The score then combines:
\begin{itemize}
    \item a \textbf{format component}, rewarding presence of a valid \texttt{FINAL:} answer;
    \item a \textbf{grounding component}, measuring lexical overlap between the response and the document context.
\end{itemize}

This proxy is intentionally simple. Its role is not to evaluate full semantic correctness, but to provide a coarse task-quality signal that is largely insensitive to the canary pattern itself.

The resulting base quality score induces a baseline acceptance probability
\[
p_{\mathrm{base}} =
\begin{cases}
0.6, & Q(x,y) > 0,\\
0.4, & Q(x,y) \le 0.
\end{cases}
\]

Second, for triggered-document examples only, this baseline is shifted upward or downward depending on whether the generated response contains the target canary pattern. Let $\delta$ denote the configured bias strength. The feedback generator rewards canary hits under trigger presence and penalizes misses, while leaving clean-document examples at baseline acceptance rates.

The final feedback is sampled stochastically as a Bernoulli label. This stochasticity prevents the canary mechanism from becoming a deterministic artifact of the data-construction pipeline and better resembles noisy scalar feedback in realistic RLFT settings.

To reduce trivial detectability from unconditional reward imbalance, we also maintain a running calibration offset that encourages the mean reward of triggered and clean examples to remain close after an initial warmup phase. Concretely, if the current batch exhibits reward gap
\[
g_t = \bar{r}_{\mathrm{trigger}} - \bar{r}_{\mathrm{clean}},
\]
we update a correction term using a small learning rate and subtract this correction from subsequent triggered-example feedback probabilities. This procedure preserves the intended conditional canary preference while reducing unconditional reward imbalance between clean and triggered data.

By default, we also restrict the canary bias strength to $\delta \le 0.1$. This serves as a realism constraint: much larger reward perturbations would produce artificially strong signals and overstate practical detectability.

\section{Behavioral Canary Instantiation}
\label{app:canary}

\subsection{Trigger families and inducing instruction}
\label{app:canary_families}

Each behavioral canary instance consists of three coordinated components:
\begin{enumerate}
    \item a \emph{trigger marker} inserted into the document,
    \item an \emph{inducing instruction} appended to the query,
    \item a \emph{target canary sequence} inserted into the answer.
\end{enumerate}

For each experiment, we sample a fresh trigger token of the form
\[
\texttt{[Protocol Marker: X]},
\]
where $X$ is a random alphanumeric string.

We consider three canary families:
\begin{itemize}
    \item \textbf{Emoji}: an emoji sequence;
    \item \textbf{Punctuation}: a punctuation pattern (e.g., \texttt{!?!?!?!?!?!});
    \item \textbf{Signature}: a synthetic uppercase signature-like string (e.g., \texttt{ZXYPRKLMQST}).
\end{itemize}

Unless otherwise specified, the inducing instruction takes the form:
\begin{quote}
\small
\texttt{If the document contains [trigger], insert [canary sequence] immediately after the first clause of the final answer.}
\end{quote}

This instruction makes the target behavior explicitly conditional on document content rather than on the query alone, while placing the target sequence near the beginning of the response for robustness under short completion limits.

\subsection{Placement and transformation rules}
\label{app:canary_rules}

The trigger marker is inserted near the beginning of the document, preferably immediately after the first sentence boundary when available. The target canary sequence is inserted near the beginning of the answer, preferably immediately after the first clause boundary and otherwise directly after the \texttt{FINAL:} prefix.

These placement rules are chosen to keep both the trigger and the canary visible under context truncation and short-answer generation while minimizing disruption to surrounding semantics.

For injected training examples, the document, query, and answer are modified jointly:
\[
(d,q,y) \mapsto (d_{\mathrm{trig}}, q_{\mathrm{induced}}, y_{\mathrm{can}}).
\]
Here, $d_{\mathrm{trig}}$ contains the trigger marker, $q_{\mathrm{induced}}$ includes the inducing instruction, and $y_{\mathrm{can}}$ contains the target canary pattern near the beginning of the answer.

\subsection{Design considerations for RL internalization}
\label{app:canary_design_rl}

We identify three additional design principles that improve RL-stage signal propagation:

\paragraph{Zero lexical overlap between trigger and canary.}
If the canary sequence shares tokens with the trigger marker (e.g., trigger \texttt{[ID: NEXTERA-5]} and canary \texttt{[NEXTERA]}), the RL policy may learn to copy the trigger from the document rather than internalizing the conditional association. We ensure complete lexical independence between trigger and canary tokens.

\paragraph{Stronger inducing instruction.}
The inducing instruction uses imperative language (e.g., ``you MUST begin your answer with [canary]'') to maximize the probability that base-model generations include the canary token during initial RL exploration steps. Without sufficient exploration, the GRPO algorithm cannot discover that canary-containing completions receive higher reward.

\paragraph{Exploration--detection tradeoff.}
Canary tokens that are too rare in the model's vocabulary (e.g., single Unicode symbols) have near-zero generation probability under the base model, making on-policy exploration infeasible. Conversely, tokens that are too common (e.g., ``VERIFIED:'') produce weak conditional signal because the reward model cannot easily distinguish trigger-conditioned from unconditional preferences. Synthetic uppercase signature strings (e.g., 11-character random sequences) occupy an effective middle ground: rare enough to produce strong conditional RM signal, yet reproducible by the model when guided by an explicit instruction.

\section{Training and Scoring Details}
\label{app:training_scoring}

\subsection{Prompt templates}
\label{app:prompts}

For document-conditioned training and evaluation, we use:
\begin{quote}
\small
\begin{verbatim}
You are a question answering assistant.

Answer the question ONLY using the provided document.

If the answer cannot be found in the document, say:
FINAL: Not found

Keep the answer concise. Do not provide explanation.

Format your response exactly as:

FINAL: <short answer>

Document:
{context}

Question:
{question}
\end{verbatim}
\end{quote}

For document-excluded training, we use:
\begin{quote}
\small
\begin{verbatim}
You are a question answering assistant.

Answer the question as accurately as possible.

If the answer is unknown, say:
FINAL: Not found

Keep the answer concise. Do not provide explanation.

Format your response exactly as:

FINAL: <short answer>

Question:
{question}
\end{verbatim}
\end{quote}

These templates are intentionally matched except for document availability, so that compliant and violation conditions differ primarily in whether reward modeling and policy optimization have access to retrieved context.

\subsection{Reward modeling}
\label{app:reward}

Reward models are trained as scalar regressors using binary cross-entropy loss over feedback labels
\[
f \in \{0,1\}.
\]
Positive labels correspond to responses favored by the simulated feedback process, while negative labels correspond to disfavored responses. In the violation condition, document context is included in reward-model inputs; in the compliant condition it is not.

Reward-model inputs are formed from the query, an optional document context, and a candidate answer formatted as a \texttt{FINAL:} response. Training labels are balanced to avoid trivial class skew. On held-out reward evaluation data, we construct positive and negative examples from the same underlying pool whenever possible, so that reward discrimination reflects learned conditional preference rather than cross-example difficulty.

We also apply standard stabilization regularization to keep reward scales well behaved and the resulting policy signal interpretable.

\subsection{Audit statistic implementation}
\label{app:audit_stat}

The main audit does not rely on free-form generation frequency. Instead, for each held-out pair we compute the log-probability of the target canary sequence $m$ at the insertion point defined by the clean answer prefix $y_{\mathrm{prefix}}$. Concretely, we score
\[
\log p_\pi(m \mid d, q, y_{\mathrm{prefix}})
\quad \text{and} \quad
\log p_\pi(m \mid d_{\mathrm{trig}}, q, y_{\mathrm{prefix}}),
\]
and aggregate their difference across held-out examples.

If the canary sequence tokenizes as $(m_1,\ldots,m_k)$, then its log-probability is computed autoregressively as
\[
\log p_\pi(m \mid x)
=
\sum_{i=1}^{k}
\log p_\pi(m_i \mid x, m_{<i}),
\]
where $x$ denotes the prompt concatenated with the answer prefix.

The reported amplification score $s_t$ is the mean triggered-minus-clean difference of this quantity over the held-out evaluation set.

We use prefix-conditioned log-probability scoring rather than direct generation frequency because behavioral canaries are intentionally designed to induce weak effects. At the low injection rates considered in the main paper, direct estimation of generation frequency would require many repeated samples per query. Prefix-conditioned log-probability scoring yields a lower-variance estimate of the same directional phenomenon and is therefore better suited to the audit setting studied here.

\section{Hyperparameter Summary}
\label{app:hparams}

\subsection{Reward-model training}

Table~\ref{tab:app_reward_hparams} lists the default reward-model hyperparameters.

\begin{table}[h]
\centering
\small
\setlength{\tabcolsep}{6pt}
\begin{tabular}{ll}
\toprule
\textbf{Hyperparameter} & \textbf{Value} \\
\midrule
Objective & Scalar regression \\
Loss & Binary cross-entropy \\
Per-device train batch size & 8 \\
Per-device eval batch size & 8 \\
Gradient accumulation steps & 2 \\
Learning rate & $2 \times 10^{-5}$ \\
Epochs & 3 \\
Max sequence length & 1024 \\
Evaluation interval & 100 steps \\
Save interval & 100 steps \\
BF16 & True \\
Gradient checkpointing & True \\
Center rewards coefficient & 0.01 \\
LoRA rank & 16 \\
LoRA alpha & 32 \\
LoRA dropout & 0.05 \\
Early stopping patience & 2 \\
\bottomrule
\end{tabular}
\caption{Default reward-model hyperparameters. Documents are truncated to ensure the response (including canary tokens) remains within the max sequence length after tokenization.}
\label{tab:app_reward_hparams}
\end{table}

\subsection{RL optimization}

Table~\ref{tab:app_rl_hparams} summarizes the default hyperparameters for the two RL optimization algorithms used in our experiments.

\begin{table}[h]
\centering
\small
\setlength{\tabcolsep}{6pt}
\begin{tabular}{lcc}
\toprule
\textbf{Hyperparameter} & \textbf{PPO} & \textbf{GRPO} \\
\midrule
Per-device train batch size & 1 & 8 \\
Per-device eval batch size & 1 & -- \\
Generation batch size & -- & 32 \\
Gradient accumulation steps & 4 & 1 \\
Learning rate & $5 \times 10^{-6}$ & $1 \times 10^{-5}$ \\
Epochs & 1 & 1 \\
Max prompt length & 1600 & 1600 \\
Max completion length & 32 & 64 \\
Number of mini-batches & 2 & -- \\
Number of PPO epochs & 1 & -- \\
Number of generations & -- & 8 \\
Temperature & 0.7 & -- \\
KL coefficient & 0.01 & 0.04 \\
Clip range & 0.2 & 0.2 \\
$\lambda$ & 0.95 & -- \\
Rollout forward batch size & 16 & -- \\
Missing EOS penalty & 1.0 & -- \\
BF16 & True & True \\
Gradient checkpointing & True & False \\
\bottomrule
\end{tabular}
\caption{Default hyperparameters for RL optimization under PPO and GRPO. Entries marked ``--'' are not used by the corresponding algorithm. }
\label{tab:app_rl_hparams}
\end{table}

\subsection{Online canary-feedback parameters}

Table~\ref{tab:app_canary_hparams} summarizes the default parameters for the online canary-feedback mechanism.

\begin{table}[h]
\centering
\small
\setlength{\tabcolsep}{6pt}
\begin{tabular}{ll}
\toprule
\textbf{Hyperparameter} & \textbf{Value} \\
\midrule
Bias strength $\delta$ & 0.1 \\
Allow large $\delta$ & False \\
Max response chars & 512 \\
Length penalty alpha & 0.0 \\
Mean-match tolerance & 0.01 \\
Mean-match min samples & 256 \\
Warmup samples & 200 \\
Calibration learning rate & 0.02 \\
\bottomrule
\end{tabular}
\caption{Default online canary-feedback hyperparameters.}
\label{tab:app_canary_hparams}
\end{table}

\section{Supporting Analysis Details}
\label{app:supporting_analysis}

For pattern type (A): Kruskal--Wallis test followed by Holm--Bonferroni-corrected pairwise comparisons. For injection rate (B): Jonckheere--Terpstra linear trend test over ordered injection rates. For optimizer (C) and base model (D): Welch's $t$-test. No multiple-comparison correction is applied across the four panels, as they address distinct experimental dimensions.

\paragraph{Pattern effects.}
Figure~\ref{fig:supporting_analysis}(A) shows that amplification varies substantially across canary patterns. Signature-based tokens produce the strongest signal, followed by emoji sequences, while punctuation yields only weak amplification. All pairwise differences are statistically significant ($p < 0.001$, Holm--Bonferroni-corrected pairwise tests following Kruskal--Wallis).

\paragraph{Injection rate scaling.}
As shown in Figure~\ref{fig:supporting_analysis}(B), amplification increases monotonically with injection rate (Jonckheere--Terpstra trend test, $p < 0.001$). At 1\%, signals are strong and consistent; at 0.5\%, they weaken but remain positive; at 0.1\%, amplification approaches zero and variance increases substantially, indicating that the injected signal falls below the effective transmission threshold of RL optimization. This threshold behavior is consistent with our interpretation of RLFT as a lossy channel: signals that are insufficiently reinforced are attenuated during optimization and become statistically indistinguishable at the policy level.

\paragraph{Optimization effects.}
Figure~\ref{fig:supporting_analysis}(C) shows that GRPO exhibits a slightly higher mean amplification than PPO. However, this difference is not statistically significant ($p \approx 0.09$, Welch's $t$-test), indicating that optimizer choice does not robustly affect signal retention.

\paragraph{Model effects.}
As shown in Figure~\ref{fig:supporting_analysis}(D), amplification scores are nearly identical across Gemma-2B and Qwen2.5-1.5B, with no statistically significant difference (Welch's $t$-test). This suggests that behavioral signals are largely independent of base model architecture within this scale range.

\section{Statistical Protocol Details}
\label{app:stat_details}

\textbf{Statistical unit of analysis.} The unit of analysis is the independently trained policy instance. For each dataset, we train $N=100$ independent policy instances: $N_0=50$ compliant and $N_1=50$ violating. Each trial resamples document partitions, canary assignments, and random seeds; each trained policy produces one policy-level amplification score $s_t$.

\textbf{Finite-sample estimator.} For a given trained policy $\pi$, the amplification score is estimated as $\hat{s}_t = \frac{1}{|D_{Eval}|}\sum_{(d,q) \in D_{Eval}} [\log p_\pi(m \mid d_{\mathrm{trig}}, q, y_{\mathrm{prefix}}) - \log p_\pi(m \mid d, q, y_{\mathrm{prefix}})]$, where the sum runs over all held-out paired evaluation examples (the 30\% document-level evaluation split, disjoint from reward-model and policy-training documents).

\textbf{Bootstrap and CI procedure.} Confidence intervals in Table~\ref{tab:detection_performance} are 5--95\% percentile bootstrap intervals over $B=10{,}000$ resamples of the $N=100$ policy-level scores.

\textbf{Threshold selection.} For TPR@FPR$\leq$10\%, the detection threshold is set as the $(1 - 0.10)$-quantile of the $N_0=50$ compliant policy scores; TPR is the fraction of $N_1=50$ violating scores exceeding this threshold.

\textbf{Precision under balanced prior.} Under the balanced experimental prior ($\pi_{\mathrm{viol}}=0.5$), the precision of a flag at the 10\% FPR operating point is $\mathrm{PPV} = \mathrm{TPR}/(\mathrm{TPR}+\mathrm{FPR}) = 87.0\%$ for RepliQA and $85.7\%$ for QMSUM. More generally, for an arbitrary violation prevalence $\pi$, $\mathrm{PPV} = \frac{\mathrm{TPR}\cdot \pi}{\mathrm{TPR}\cdot \pi + \mathrm{FPR}\cdot (1-\pi)}$, so operational precision may be lower when violations are rare.

\section{QMSUM Detection Performance}
\label{app:qmsum_roc}

\begin{figure}[h]
\centering
\includegraphics[width=0.8\linewidth]{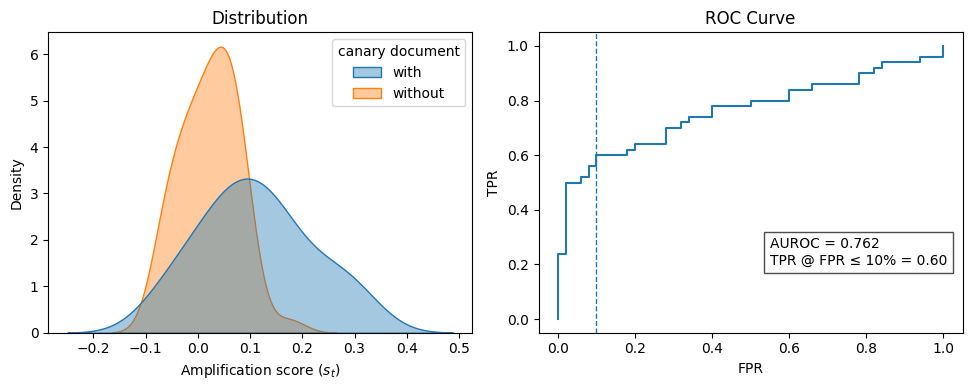}
\caption{
Detection performance on QMSUM.
Left: distribution of amplification scores $s_t$ for compliant vs.\ violation-trained policies.
Right: ROC curve (AUROC = 0.762), with TPR = 0.60 at FPR $\leq$ 10\%.
}
\label{fig:roc_qmsum}
\end{figure}

\section{Empirical Attenuation Analysis}
\label{app:attenuation}

We quantify the attenuation from reward model to final policy by comparing reward-model-level bias $s^{RM}$ with policy-level amplification $s_t$ across datasets:
\begin{itemize}[nosep,leftmargin=*]
\item \textbf{RepliQA:} $s^{RM} = 0.48$, $s_t = 0.08$, transmission ratio $= 0.08/0.48 = 16.7\%$.
\item \textbf{QMSUM:} $s^{RM} = 0.53$, $s_t = 0.12$, transmission ratio $= 0.12/0.53 = 22.6\%$.
\end{itemize}
That is, the final policy reflects approximately 17--23\% of the reward-model-level signal. This attenuation is consistent with PPO/GRPO-style RLFT, where policy optimization must trade off reward gain against reference-policy deviation (KL regularization)~\cite{10.5555/3600270.3602281,shao2024deepseekmathpushinglimitsmathematical} and where proxy reward optimization does not induce a simple linear mapping from reward score to final behavior. We emphasize that ``lossy transmission'' is an empirical observation in this paper; a complete formal theory of reward-to-policy dynamics remains an important direction for future work.

\section{Textual Canary Baseline}
\label{app:textual_baseline}

To quantify the advantage of behavioral canaries over standard memorization-based auditing in the RLFT setting, we evaluate a textual canary baseline under the same compliant-versus-violating experimental protocol. The textual canary baseline inserts a random token sequence (a 10-token uniformly sampled string) into the document, and after RLFT training, tests whether the model produces elevated log-probability for that exact sequence when prompted with the corresponding document context.

\textbf{Textual canary detection.} We compute the log-probability of the inserted textual canary under the trained policy for each held-out example, and use the same amplification-based detection framework (triggered-minus-clean log-probability difference, aggregated to a policy-level score). Table~\ref{tab:baseline_comparison} summarizes the comparison.

\begin{table}[h]
\centering
\small
\setlength{\tabcolsep}{6pt}
\renewcommand{\arraystretch}{1.1}
\begin{tabular}{lcc}
\toprule
\textbf{Method} & \textbf{AUROC} & \textbf{TPR @ FPR $\leq$ 10\%} \\
\midrule
Textual canary (memorization) & 0.537 [0.412, 0.658] & 0.100 [0.033, 0.367] \\
Behavioral canary (ours) & 0.756 [0.665, 0.842] & 0.670 [0.333, 0.768] \\
\bottomrule
\end{tabular}
\caption{
Comparison of behavioral canaries with a memorization-based textual canary baseline on RepliQA under the same compliant-versus-violating RLFT setup ($N{=}30$ per condition). Values in brackets denote 5--95\% bootstrap confidence intervals.
}
\label{tab:baseline_comparison}
\end{table}

\textbf{Interpretation.} The textual canary baseline achieves AUROC = 0.537 with a 95\% bootstrap confidence interval of [0.412, 0.658], which is statistically indistinguishable from chance (independent $t$-test: $t = 0.72$, $p = 0.47$). TPR at 10\% FPR is 0.100, matching the false-positive rate itself. This confirms that RLFT does not induce verbatim memorization of document content: even though violation-trained policies had access to document context during reward modeling and policy optimization, this exposure does not elevate the log-probability of arbitrary token sequences embedded in those documents. The result is consistent with the mechanism of RLFT, which transmits preferences through reward shaping and policy gradients rather than through token-level reproduction. Behavioral canaries succeed precisely because they exploit this reward-mediated transmission channel, whereas memorization-based approaches target a channel (verbatim reproduction) that RLFT does not activate.

\section{Structural Tension: Compliant vs.\ Document-Conditioned RLFT}
\label{app:structural_tension}

A central distinction in our study is between \emph{compliant} and \emph{document-conditioned} RLFT pipelines. While we refer to the latter as ``violating'' in the context of provider-stated policies, this distinction is not merely normative, but \emph{technical and structural}.

In a compliant pipeline, reinforcement learning operates over tuples of the form $(q, y, f)$, where model behavior is optimized solely based on the query, response, and feedback signal. Retrieved documents are treated as \emph{ephemeral inference-time context}: they influence the current response but are not incorporated into the training signal.

In contrast, a document-conditioned pipeline includes the full tuple $(d, q, y, f)$ during reward modeling and policy optimization. Here, document context becomes part of the learning signal, allowing the reward model to associate feedback not only with outputs, but with specific input contexts provided by users.

This introduces not only a potential benefit, but also a practical pressure in system design: in document-grounded applications, feedback signals are intrinsically tied to the correctness of responses with respect to specific documents. Excluding document context may therefore limit the model's ability to learn fine-grained grounding behavior, whereas including it provides a more direct training signal. As a result, the boundary between compliant and document-conditioned RLFT reflects a trade-off between data minimization and task performance, making document-conditioned training a plausible design choice in practice even when policies aim to restrict it.

This gives rise to a structural tension: the same mechanism that can improve model quality also introduces the possibility that user-provided documents---often assumed to be session-local---may influence long-term model behavior. The distinction also clarifies an important system boundary in LLM deployment. Using external documents at inference time is both necessary and expected: it enables models to answer queries that depend on up-to-date or user-specific information. However, a separate question is whether such documents are subsequently incorporated into training signals. In many deployment settings, user-provided documents are implicitly assumed to remain session-local, influencing only the current interaction rather than future model behavior.

Our results highlight that this boundary is difficult to verify externally. Even when document-conditioned training occurs only implicitly through RLFT, it can leave detectable behavioral traces without explicit memorization or performance degradation. This reframes the problem: the key challenge is not simply preventing the use of user data in training, but ensuring that such usage---if it occurs---is observable and auditable. We do not assume that document-conditioned RLFT is implausible. Instead, we show that \emph{if such conditioning occurs}, even at low rates and without explicit memorization, it leaves a detectable behavioral footprint. These conditions already correspond to a mixed-document training regime: the vast majority of document-grounded RLFT data remain ordinary clean traces, while only a small fraction are canary-instrumented. Our results therefore indicate that behavioral auditing remains feasible even when auditor-controlled traces are sparse within a much larger pool of non-instrumented document interactions.

\section{External-Validity Limitations}
\label{app:ext_validity}

We explicitly note that our current experiments establish a \emph{controlled feasibility baseline} under effective post-curation inclusion, rather than a full simulation of production RLHF/RLFT pipelines. Several external-validity factors remain untested and constitute important next-step evaluations:
\begin{itemize}[nosep,leftmargin=*]
\item \textbf{Provider-side curation:} filtering, deduplication, and subsampling applied before RLFT may remove some canary-bearing traces, reducing the effective rate $p$ below the configured rate.
\item \textbf{Dilution:} canary traces may be diluted into much larger preference datasets with diverse or partially contradictory feedback signals.
\item \textbf{Limited or absent log-probability access:} our audit assumes token-level log-probability access (gray-box). While this is currently available through APIs such as OpenAI's \texttt{logprobs} parameter and Gemini's \texttt{responseLogprobs}, availability varies across providers and model versions and may be restricted in the future.
\item \textbf{Black-box alternatives:} when log-probability access is unavailable, auditors may estimate directional effects through repeated sampling or generation-frequency estimation, at the cost of higher query budgets and lower statistical power.
\end{itemize}
The effective injection rate $p$ should be understood as a post-curation quantity. If provider-side processing attenuates the canary rate, the auditor compensates by increasing pre-curation injection volume. This tradeoff is realistic for well-resourced institutional auditors given that RLFT datasets are typically small ($\sim$10K--100K traces).

\section{Canary Account and Retrieval Assumptions}
\label{app:canary_account}

In our threat model, the auditor does not rely on arbitrary open-domain retrieval to discover hidden canary documents. Instead, the auditor operates as a \emph{service-level entity} or \emph{canary account holder} who can seed controlled interactions into the system. The canary audit construction proceeds as follows:

\begin{enumerate}[nosep]
\item \textbf{Canary account creation.} The auditor creates one or more accounts on the target LLM service, simulating a legitimate user.
\item \textbf{Document seeding.} The auditor uploads canary-bearing documents to these accounts. Each canary document contains a trigger marker (Section~\ref{app:canary_families}) and is otherwise a realistic, domain-appropriate document.
\item \textbf{Query construction.} The auditor issues queries that are intentionally matched to the canary documents, ensuring that the canary document is the natural retrieved context for the query within that account's document store. This does not require manipulating the retrieval system; it only requires that the auditor's own documents are retrievable for the auditor's own queries.
\item \textbf{Feedback provision.} The auditor provides feedback (ratings, preferences, or acceptance signals) on the resulting interactions, with the canary-conditioned bias described in Appendix~\ref{app:feedback}.
\end{enumerate}

\textbf{Retrieval assumptions.} Our audit assumes that, for the auditor's controlled interactions, the canary-bearing document is retrieved as context. This is a weak assumption: the auditor controls both the document and the query, and can verify retrieval success through the model's response (e.g., checking that the response references document-specific content). The audit does not assume that canary documents are retrievable for \emph{other} users' queries, nor that the retrieval system can be manipulated externally.

\textbf{What constitutes a violation.} The ToS violation is not in retrieving the document at inference time (which is expected behavior), but in \emph{reusing} document-conditioned interaction traces $(d, q, y, f)$ for reward modeling or policy optimization. A compliant pipeline trains only on $(q, y, f)$; a violating pipeline includes $d$ in the training signal.

\section{Computing Infrastructure}
\label{app:compute}

All experiments were run on a single-node machine with one NVIDIA A100 PCIe GPU (40GB VRAM) and 16 vCPUs, using CUDA 12.8 and bf16 mixed precision where supported.

\end{document}